\documentclass[prd, preprint,amscd,amsmath,amssymb,verbatim,showpacs,nofootinbib]{revtex4}
\usepackage{graphicx} 
\usepackage{epstopdf} 
\usepackage{tensor}
\usepackage{amsfonts}
\usepackage{bm}

\newcommand{\be}{\begin{equation}}
\newcommand{\ee}{\end{equation}}
\newcommand{\bea}{\begin{eqnarray}}
\newcommand{\eea}{\end{eqnarray}}
\newcommand{\beaa}{\begin{eqnarray*}} 
\newcommand{\eeaa}{\end{eqnarray*}}

\newcommand{\dis}{\displaystyle} 
\newcommand{\bsube}{\begin{subequations}}
\newcommand{\esube}{\end{subequations}}
\begin{document}

\title{Combined gravitational and electromagnetic self-force on charged particles in 
electrovac spacetimes}
\author{Thomas M. Linz}
\email{tmlinz@uwm.edu}
\affiliation{ Leonard E Parker Center for Gravitation, Cosmology and Astrophysics, Department of Physics,
University of Wisconsin--Milwaukee, P.O. Box 413, Milwaukee, Wisconsin 53201,
USA}

\author{John L. Friedman}
\email{friedman@uwm.edu}
\affiliation{ Leonard E Parker Center for Gravitation, Cosmology and Astrophysics, Department of Physics,
University of Wisconsin--Milwaukee, P.O. Box 413, Milwaukee, Wisconsin 53201,
USA}

\author{Alan G. Wiseman}
\email{agw@gravity.phys.uwm.edu}
\affiliation{ Leonard E Parker Center for Gravitation, Cosmology and Astrophysics, Department of Physics,
University of Wisconsin--Milwaukee, P.O. Box 413, Milwaukee, Wisconsin 53201,
USA}

\date{December 11, 2013}
\pacs{04.30.Db, 04.25.Nx, 04.70.Bw}

\begin{abstract}

We consider the self-force on a charged particle moving in a curved spacetime 
with a background electromagnetic field, extending previous studies to 
situations in which gravitational and electromagnetic perturbations are 
comparable. The formal expression $f^{ret}_\alpha$ 
for the self-force on a particle, written in terms of the retarded perturbed 
fields, is divergent, and a renormalization is needed to find the particle's 
acceleration at linear order in its mass $m$ and charge $e$.  We assume that, 
as in previous work in a Lorenz gauge, the renormalization for accelerated motion 
comprises an angle average and mass renormalization.  Using the short distance 
expansion of the perturbed electromagnetic and gravitational fields, 
we show that the renormalization is equivalent to that obtained from a mode sum regularization 
in which one subtracts from the expression for the self-force in terms of the 
retarded fields a singular part field comprising only the leading and 
subleading terms in the mode sum.  The most striking part of 
our result, arising from a remarkable cancellation, is that the 
renormalization involves no mixing of electromagnetic and gravitational 
fields. In particular, the renormalized mass is obtained by subtracting 
(1) the purely electromagnetic contribution from a point charge moving along 
an accelerated trajectory and (2) the purely gravitational contribution 
from a point mass moving along the same trajectory.  In a mode-sum regularization, 
the same cancellation implies that the required regularization parameters 
are sums of their purely electromagnetic and gravitational values.  

\end{abstract}

\maketitle 

\section{Introduction}
\label{Introduction}

The last fifteen years have seen a number of papers dealing with the self-force on a 
small mass $m$, modeled as a particle moving on a background spacetime, and 
on the analogous problem for a charge $e$ or scalar charge $q$ (see reviews by 
Barack \cite{barack09} and Poisson {\em et al.} \cite{Poisson}).  The work on 
motion of a small mass has been restricted to uncharged particles whose motion at zeroth 
order in $m$ is a geodesic of an unperturbed vacuum spacetime; corresponding work 
on charged particles has assumed a negligible contribution to the self-force from 
the perturbed gravitational field.  The primary motivation has been to 
study extreme mass-ratio inspiral (EMRI), the inspiral of stellar-size black 
holes or neutron stars orbiting supermassive galactic black holes, with 
the electromagnetic and scalar studies serving as toy problems.  There has, 
however, been recent interest in whether self-force plays a fundamental 
role in enforcing cosmic censorship by preventing one from overcharging (or 
overspinning) a near-extreme black hole \cite{Hubeny,ist,bck,zvph}.  In this 
context, one would like to analyze scenarios in which gravitational and 
electromagnetic perturbations have comparable magnitude. 

A charged particle moving on a smooth background spacetime $M,g_{\alpha\beta}$ with 
electromagnetic field $F_{\alpha\beta}$ has trajectory $z(\tau)$ satisfying the 
Lorentz force law 
\be
   {m} a_\alpha =  e F_{\alpha\beta} u^\beta, 
\label{BackGround0}\ee
For a smooth perturbation $g_{\alpha\beta} + h_{\alpha\beta}$, $F_{\alpha\beta}+\delta F_{\alpha\beta}$ 
of the geometry and electromagnetic field, the 4-velocity $\bar u^\alpha$ of the perturbed 
trajectory satisfies 
\be
{m} \bar u^\beta \nabla_\beta \bar u_\alpha - e F_{\alpha\beta}\bar u^\beta 
	= \delta F_{\alpha\beta}\ u^\beta - q_\alpha^\delta(\nabla_\beta h_{\gamma\delta}-\frac12\nabla_\delta h_{\beta\gamma})u^\beta u^\gamma := f_\alpha = f^{EM}_\alpha+ f^{GR}_\alpha,
\label{sf_formal}\ee
with $\bar u^\alpha$ normalized by the background metric, 
\be
   g_{\alpha\beta} \bar u^\alpha \bar u^\beta = -1.
\ee
When the perturbation is due to the retarded fields of the particle itself, the 
formal expression (\ref{sf_formal}) for the self-force, with 
$f^{ret}_\alpha = f^\alpha[h^{ret},\delta F^{ret}]$, diverges at the particle.  
Mino {\em et al.} \cite{MST} and Quinn and Wald \cite{QW} obtained 
equivalent prescriptions for renormaliz‌ing $f^{ret}_\alpha$, following work by 
DeWitt and Brehme \cite{DewittBrehme} (corrected by Hobbs \cite{hobbs}) on a 
charged particle moving in a vacuum background spacetime.%
\footnote{The most recent and rigorous work justifying this MiSaTaQuWa renormalization
uses variants of matched asymptotic expansion to find the motion of small bodies in 
the limit where the size of the body and its mass (or charge) simultaneously shrink to 
zero.  See Gralla {\em et al.} \cite{gw08,ghw09} (with a formal proof for an electromagnetic charge), 
Pound \cite{pound10}, and Poisson {\em et al.} \cite{Poisson}, 
who also review the history and give a comprehensive bibliography.}  

When the unperturbed motion is geodesic, the renormalized self-force at a point $z$ of the 
particle's trajectory can be obtained as the $\rho\rightarrow 0$ limit of 
an angle average of $f^{ret}_\alpha$ over a sphere $S_\rho$ of geodesic distance 
$\rho$ from $z$ \cite{gralla11}.  Explicitly,  
\be
f^{ren}_{\alpha}(z)  
   =\lim_{\rho\rightarrow 0} \langle f^{ret}_{\alpha}\rangle_{\rho} 
	= \lim_{\rho\rightarrow 0} \int_{S_\rho} d\Omega f^{ret}_{\alpha},
\label{SFansatz0}
\ee
where the components $f^{ret}_\alpha$ are given in Riemann normal coordinates (RNCs) 
centered at $z$. (Equivalently, the average is taken in the tangent space 
at $z$ with $f^{ret}_\alpha$ pulled back by the exponential map.) 
When the trajectory is accelerated, we show in a previous paper \cite{LinzFriedmanWiseman} (henceforth {\em Paper I}) that the angle average leaves a term proportional 
to $a_\alpha/\rho$, which can be regarded as a renormalization of the mass.  
The renormalized self-force on an electromagnetic or scalar charge moving on an 
accelerated trajectory has the form 
\be
f^{ren}_{\alpha}
   = \lim_{\rho\rightarrow 0}\langle f^{ret}_{\alpha}\rangle_\rho
		-m^{sing}(\rho) a_{\alpha},
\label{SFansatz}
\ee
with $m^{sing}(\rho)\propto \rho^{-1}$.  
For the more general situation we consider here, with electromagnetic and gravitational 
perturbations each contributing to the self-force, we again assume 
that $f^{ren}_{\alpha}$ is given by Eq.~(\ref{SFansatz}).

We find that $m^{sing}$ is a sum $m^{sing} = m^{GR}+ m^{EM}$ of gravitational and electromagnetic 
parts.  $m^{GR}$ is proportional to ${m}^2$ and comes solely from the gravitational 
perturbation associated with the accelerated motion of a mass $m$; 
$m^{EM}$ is proportional to $e^2$ and comes solely from the electromagnetic perturbation 
associated with the motion of a charge $e$.  Contributions to $f^{sing}_\alpha$ 
from terms proportional to $em$ are equal and opposite and give no net 
mixed contribution to $m^{sing}$.   
  
We show that this renormalization is equivalent to performing an angle average after 
subtracting a singular part of $f^{ret}_\alpha$ associated with the short-distance expansion 
of the perturbations $h_{\alpha\beta}$ and $\delta A_\alpha$ to subleading order in $\epsilon$.
An independent calculation by Zimmerman and Poisson finds the Detweiler-Whiting forms 
of the singular fields $h^S_{\alpha\beta}$ and $\delta^S A_\alpha$ and the 
corresponding form $f^S_\alpha$ of the singular part of the expression for the 
self force, and our renormalizations agree.   

Most explicit calculations of the self-force on particles moving in
Kerr or Schwarzschild geometries, however, have used a mode-sum form 
of the renormalization introduced by Barack and Ori \cite{BO00,BO1}, with 
early development and first applications by them, by Mino {\em et al.}, and by 
Burko \cite{BMNOS,BarackBurko,Burko} (see Refs.~\cite{barack09,Poisson} for reviews 
of later work). Here one expresses $f^{sing}_\alpha$ and $f^{ret}_\alpha$ as sums of angular 
harmonics on a sphere through the particle, writing the renormalized self-force as 
the convergent sum $\dis{ \sum_{\ell=0}^{\infty}}(f^{ret,\ell}_\alpha - f^{sing,\ell}_\alpha)$,
with $f^{ret,\ell}_\alpha$ and $f^{sing,\ell}_\alpha$ each a sum over $\texttt{m}$ of its $\ell, \texttt{m}$ harmonics.  
Paper I extended to accelerated motion a fundamental feature of this mode sum in 
a Lorenz or smoothly related gauge: Only the leading 
and subleading terms in $\ell^{-1}$ give nonzero contributions to the singular 
expression for the self-force.  For a point particle with scalar or electromagnetic charge, 
and for a point mass, $f^{sing,\ell}_\alpha$ has the form 
\be
    f^{sing,\ell\pm}_\alpha = \pm A_\alpha L + B_\alpha,
\label{eq:fsingAB}\ee 
where $L=\ell+1/2$, $A_\alpha$ and $B_\alpha$ are independent of $\ell$,
and the sign $\pm$ refers to a limit of the direction-dependent 
singular expression taken as one approaches the sphere through the particle 
from the outside or inside.   We show here that the same form holds for a 
charged particle moving in an electrovac spacetime.  The coupling of 
electromagnetic and gravitational perturbations in the Einstein-Maxwell system 
means that the renormalized self-force has mixed contributions proportional 
to $em$, but no such mixed terms arise in the renormalization:  
The cancellation mentioned above implies that 
{\em the regularization parameters $A_\alpha$ and $B_\alpha$ are just the 
sums of their values for purely gravitational and purely electromagnetic 
contributions} to the terms arising in mode-sum regularization of an
accelerated particle $e,m$.  (Higher order regularization parameters 
are useful for convergence, and they presumably do involve mixed terms, 
but they multiply vanishing sums.)  

The plan of the paper is as follows.  In Sect.~\ref{SectionII}, we find explicit expansions 
in RNCs for the electromagnetic and gravitational perturbations produced by a particle of mass 
$m$ and charge $e$ moving in an electrovac background spacetime.  These expansions can be 
identified with the singular parts $h^{sing}_{\alpha\beta}$ and 
$\delta A_\alpha $ of the perturbations and provide an expression for the 
$f^{sing}_\alpha$, again to subleading order.  We obtain the cancellation of 
mixed gravitational and electromagnetic terms (terms proportional to $em$) mentioned 
above, and write the explicit expression for the mass renormalization.  
In Sect.~(\ref{SectionIII}), we obtain the regularization parameters $A_\alpha$ 
and $B_\alpha$ required for mode-sum regularization of the self-force.  
They are the simply the sum of terms previously obtained in Paper I for 
accelerated charges and accelerated masses, respectively.  
Finally, using forms for the perturbed gravitational and scalar field 
of particle with scalar charge moving in a scalarvac spactime provided by 
Poisson and Zimmerman \cite{pz14}, we show that
in the renormalization of the self-force, the contributions of gravitational 
and electromagnetic fields are similarly decoupled.

\section{Self-force in electrovac spacetimes}

\label{SectionII} 

We consider a point particle of mass $m$ and charge $e$ moving with trajectory $z(\tau)$ in a smooth electrovac spacetime, $(M,g_{\alpha\beta}, F_{\alpha\beta})$, with $F_{\alpha\beta}$ a sourcefree electromagnetic field.  The metric $g_{\alpha\beta}$ of the background spacetime then has as its source the stress-energy tensor of $F_{\alpha\beta}$, 
\be
G_{\alpha\beta}=8\pi T_{\alpha\beta}=2\left(F_{\alpha\mu}F\indices{_{\beta}^{\mu}}-\frac{1}{4}g_{\alpha\beta}F^{\mu\nu}F_{\mu\nu}\right),
\label{BackGroundG}
\ee
where $F_{\alpha\beta}$ satisfies 
\be
\nabla_{\beta}F^{\alpha\beta}=0,\qquad \nabla_{[\alpha}F_{\beta\delta]}=0.
\label{BackGroundF}
\ee
We are interested in the self-force per unit mass on the particle at linear order in $m$ and $e$.
To make this precise, one could consider a family of solutions $g_{\alpha\beta}({m},e), 
F_{\alpha\beta}({m},e)$ whose source for nonzero $m$ and $e$ is a body of finite extent, 
where $e/m$ has a finite limit as $m\rightarrow 0$ and where the characteristic 
spatial length of the body is, like $e$, linear in $m$ for small $m$. 
At ${m}=0$, the spacetime is the electrovac background, and the ${m}\rightarrow 0$ 
limit of the family of trajectories is given by the Lorentz force law of that background, 
\be
   a_\alpha = \frac e{m} F_{\alpha\beta} u^\beta, 
\label{BackGrounda}\ee
where $u^\alpha$ is the particle's velocity, and $a^\alpha = u^\beta\nabla_\beta u^\alpha$ is its 
acceleration relative to the background geometry, and $\nabla_\alpha$ is the covariant derivative of the background metric.  The self-force arises from the perturbations 
in the gravitational and electromagnetic fields due to the body.  
We denote by $\delta Q$ the linear perturbation in a quantity $Q({m},e)$, 
\be
\delta Q := 
 \left.{m}\ \frac\partial{\partial{m}}Q({m},e)\right|_{({m},e)=(0,0)} 
+ \left. e\ \frac\partial{\partial e} Q({m},e) \right|_{ ({m},e)=(0,0)}.  
\label{deltaQ}\ee
Then $Q({m},e) = Q + \delta Q + O({m}^2, em, e^2)$, where $Q\equiv Q(0,0)$.
The perturbations $h_{\alpha\beta}= \delta g_{\alpha\beta}$ and $\delta F_{\alpha\beta}$ are the linearized gravitational 
and electromagnetic fields of a point particle with trajectory satisfying 
Eq.~(\ref{BackGrounda}).  In the problems that motivate this approximation, 
the background spacetime is nonradiative and the perturbations are the retarded fields 
$h^{\rm ret}_{\alpha\beta}$ and $\delta F^{\rm ret}_{\alpha\beta}$ of the particle,
but the renormalization procedure is unrelated to these restrictions.  
 
In the remainder of the paper, as in the previous paragraph, 
the symbols $g_{\alpha\beta}$ and $F_{\alpha\beta}$ will refer to the background metric 
and electromagnetic field, and all indices will be raised and lowered by the background 
metric.  
  

    We assume that, to linear order in the perturbed fields, the trajectory $z(\tau)$ of the 
particle satisfies the renormalized Lorentz-force law equation associated with 
the perturbed metric $g_{\alpha\beta}+h_{\alpha\beta}$ and electromagnetic field 
$F_{\alpha\beta}+\delta F_{\alpha\beta}$,     
\be
{m} u^\beta({m},e)  \nabla_\beta u_\alpha({m},e) 
	= e u^\beta({m},e )F_{\alpha\beta} + f_\alpha^{ren}
	  + O({m}^2, em, e^2).
\ee
where $f_\alpha^{ren}$ is obtained from the formal expression (\ref{sf_formal}) for the self-force 
by angle average and mass renormalization, as in Eq.~(\ref{SFansatz}).   
 
We will show that the renormalization of Eq.~(\ref{SFansatz}) is equivalent to 
separate renormalization of the electromagnetic and gravitational contributions 
to the self-force $f_\alpha$. It will then follow that in the mode-sum 
renormalization, there is no mixing of gravitational and electromagnetic 
parts:  The renormalization is equivalent to subtracting (1) a singular expression 
$f^{sing}_\alpha = f^{EM,sing}_\alpha + f^{GR,sing}_\alpha$, 
where $f^{EM,sing}_\alpha$ is the purely electromagnetic contribution 
from a point charge moving along an accelerated trajectory (with no 
perturbed gravitational field); and $f^{GR,sing}_\alpha$ is the 
the purely gravitational contribution from a point mass moving along 
the same trajectory that would arise if there were no perturbed 
electromagnetic field.

We consider the field in a convex normal neighborhood $C$ of the event $z(0)$, 
denote by $x$ any point of $C$ and by $\epsilon$ the length of the unique 
geodesic from $z(0)$ to $x$. 
\begin{figure}[h!]
\begin{center}
\includegraphics[width=3in]{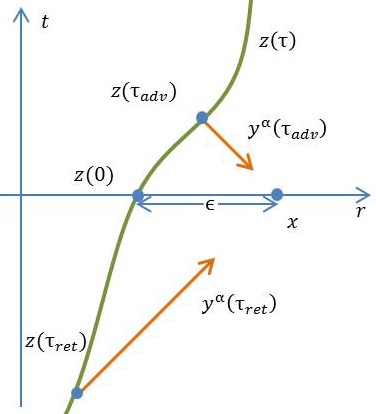}
\caption{The particle trajectory $z(\tau)$.  Two null vectors  
$y^\alpha(\tau_{ret})$ and $y_{\alpha}(\tau_{adv})$ are tangent 
to future- and past-directed null geodesics from points along 
the trajectory to a field point $x$. A geodesic from $z(0)$ 
to $x$ has length $\epsilon$.}
\label{Fig1}
\end{center}
\end{figure}
We choose $\tau=0$ at the position of the particle where we renormalize.

We work in a Lorenz gauge for each field: Introducing the trace-reversed metric perturbation 
\be
 \gamma_{\alpha\beta}=h_{\alpha\beta}-\frac12 g_{\alpha\beta} h\indices{^\delta_\delta}
\ee
and a vector potential $\delta A_\alpha$ for which 
$\delta F_{\alpha\beta} = \nabla_\alpha \delta A_\beta - \nabla_\beta \delta A_\alpha$, 
we have 
\be
  \nabla^\beta \gamma_{\alpha\beta} = 0, \qquad \nabla^\beta \delta A_\beta = 0.
\ee
In this gauge, the perturbed Einstein equation, $\delta G_{\alpha\beta}=8\pi \delta T_{\alpha\beta}$, has the form
\bea
-2\delta G_{\alpha\beta}&=&\square\gamma_{\alpha\beta}+2\Omega\indices{_{\alpha}^{\gamma}_{\beta}^{\delta}}\gamma_{\gamma\delta}\nonumber\\
&=&-16\pi{m}\int u_{\alpha}u_{\beta}\delta^{(4)}(x,z(\tau))d\tau-8\left(F\indices{_{(\alpha}^{\delta}}\delta\indices{_{\beta)}^{\gamma}}-\frac{1}{4}g_{\alpha\beta}F^{\gamma\delta}\right)\delta F_{\gamma\delta}\nonumber\\
&+&\left[4F\indices{_{\alpha}^{\gamma}}F\indices{_{\beta}^{\delta}}-2F_{\alpha\epsilon}F\indices{_{\beta}^{\epsilon}}g^{\gamma\delta}-g_{\alpha\beta}F\indices{^{\gamma}_{\epsilon}}F^{\delta\epsilon}+F_{\epsilon\gamma}F^{\epsilon\gamma}\left(\delta^{\gamma}_{\alpha}\delta^{\delta}_{\beta}+\frac{1}{2}g_{\alpha\beta}g^{\gamma\delta}\right)
\right]\gamma_{\gamma\delta},\nonumber\\
\label{PerturbG}
\eea
where $\Box = \nabla_\alpha \nabla^\alpha$ and 
\be
\Omega\indices{_{\alpha}^{\gamma}_{\beta}^{\delta}} := R\indices{_{(\alpha}^{\gamma}_{\beta)}^{\delta}}-R_{(\alpha}^{\gamma}\delta_{\beta)}^{\delta}-\frac{1}{2}g_{\alpha\beta}R^{\gamma\delta}+\frac{1}{2}R\delta_{(\alpha}^{\gamma}\delta_{\beta)}^{\delta}.
\label{Omega}
\ee
To make the notation more concise, we combine the last line of Eq.~(\ref{PerturbG}) with the 
term $2\Omega\indices{_{\alpha}^{\gamma}_{\beta}^{\delta}}\gamma_{\gamma\delta}$ 
to write
\be
\square\gamma_{\alpha\beta}+2\hat{\Omega}\indices{_{\alpha}^{\gamma}_{\beta}^{\delta}}\gamma_{\gamma\delta}
=-16\pi{m}\int u_{\alpha}u_{\beta}\delta^{(4)}(x,z(\tau))d\tau-16\left(F\indices{_{(\alpha}^{[\delta}}\delta\indices{_{\beta)}^{\gamma]}}-\frac{1}{4}g_{\alpha\beta}F^{\gamma\delta}\right)\partial_{\gamma}\delta A_{\delta},
\label{PerturbEinstein}
\ee
with
\bea
\hat{\Omega}\indices{_{\alpha}^{\gamma}_{\beta}^{\delta}} := \Omega\indices{_{\alpha}^{\gamma}_{\beta}^{\delta}}
 - 2F\indices{_{(\alpha}^{\gamma}}F\indices{_{\beta)}^{\delta}} 
 + F\indices{_{(\beta}^{\epsilon}}F_{\alpha)\epsilon} g^{\gamma\delta}
 + g_{\alpha\beta}F\indices{^{\gamma}_{\epsilon}}F^{\delta\epsilon}
 -\frac12 F_{\epsilon\gamma}F^{\epsilon\gamma}\left(\delta^{\gamma}_{(\alpha}\delta^{\delta}_{\beta)}
 -\frac{1}{2}g_{\alpha\beta}g^{\gamma\delta}\right).\quad
\label{Omegahat}
\eea
The perturbed Maxwell equation, $\delta (\nabla_\beta F^{\alpha\beta}) = 4\pi \delta j^\alpha$, is given by 
\be
\square\delta A_{\alpha}-R_{\alpha}^{\beta}\delta A_{\beta}=-4\pi e \int u_{\alpha} \delta^{(4)}(x,z(\tau))d\tau-\nabla^{\beta}\left[\left(F\indices{^{\gamma}_{\beta}}\delta^{\delta}_{\alpha}+F\indices{_{\alpha}^{\delta}}\delta^{\gamma}_{\beta}-\frac{1}{2}g^{\gamma\delta}F_{\alpha\beta}\right)\gamma_{\gamma\delta}\right].
\label{PerturbMaxwell}
\ee

 To find the singular behavior of the perturbed fields $\gamma_{\alpha\beta}$ and 
$\delta A_{\alpha}$, we follow the formalism described in 
Paper I \cite{LinzFriedmanWiseman}.  We introduce Riemann normal coordinates 
(RNCs) $\{x^\mu\}$ with origin at $z(0)$ and find the coordinate 
expansion of the perturbed fields. 
As in the case of particles with purely electromagnetic or gravitational interactions, 
 the angle-average renormalization of Eq.~(\ref{SFansatz}) is 
equivalent to identifying and subtracting from $f^{ret}_\alpha$ a singular 
part $f^{sing,\alpha}$, for which the difference $f^{ret}_\alpha - f^{sing,\alpha}$ is continuous at the position of the particle.  The singular expression $f^{sing}_{\alpha}$
is in turn obtained from Eq.~(\ref{sf_formal}) by replacing $\gamma_{\alpha\beta}$ 
and $\delta A_\alpha$ by singular parts $\gamma^{sing}_{\alpha\beta}$ 
and $\delta A^{sing}_\alpha$ of the perturbed fields.  
As shown in Paper I,  this prescription, inspired by the work of Gralla and Wald \cite{gw08}, agrees exactly with the Detweiler-Whiting singular field \cite{DW} when we apply it to scalar, or electric charges in any spacetime and for point masses in vacuum. 
A comparison with the singular potentials found by Poisson and Zimmerman \cite{pz14} 
using the Detweiler-Whiting singular fields shows that shows that the angle-average renormalization is again equivalent to the renormalizing using the Detweiler-Whiting
prescription for the renormalized Green's functions.  


Our approach to expanding the perturbed fields is tailored to the mode-sum regularization methods pioneered by Barack and Ori \cite{BO00}. We decompose the field perturbations 
into two pieces, $\gamma_{\alpha\beta}={_{I}\gamma}_{\alpha\beta}+{_{II}\gamma}_{\alpha\beta}$ and $\delta A={_{I}A}_{\alpha}+{_{II}A}_{\alpha}$, satisfying 
\bea
\square\ {_{I}\gamma}_{\alpha\beta}+2\hat{\Omega}\indices{_{\alpha}^{\gamma}_{\beta}^{\delta}}{_{I}\gamma}_{\gamma\delta}&=&-16\pi{m}\int u_{\alpha}u_{\beta}\delta^{(4)}(x,z(\tau))d\tau
\label{PertIG},\\
\square\ {_{I}A}_{\alpha}-R_{\alpha}^{\beta}{_{I}A}_{\beta}&=&-4\pi e\int u_{\alpha}\delta^{(4)}(x,z(\tau))d\tau,
\label{PertIE}
\eea
and
\bea
\square\ {_{II}\gamma}_{\alpha\beta}+2\hat{\Omega}\indices{_{\alpha}^{\gamma}_{\beta}^{\delta}}{_{II}\gamma}_{\gamma\delta}&=&-16\Lambda\indices{_{\alpha\beta}^{\gamma\delta}}\partial_{\gamma}\delta A_{\delta},
\label{PertIIG}\\
\square\ {_{II}A}_{\alpha}-R_{\alpha}^{\beta}{_{II}A}_{\beta}&=&-2\nabla^{\beta}\left[\Lambda\indices{^{\gamma\delta}_{\alpha\beta}}\gamma_{\gamma\delta}\right],
\label{PertIIE}
\eea
where
\bea
\Lambda\indices{_{\alpha\beta}^{\gamma\delta}}&=&F\indices{_{(\alpha}^{[\delta}}\delta\indices{_{\beta)}^{\gamma]}}-\frac{1}{4}g_{\alpha\beta}F^{\gamma\delta}.
\label{Lambda}
\eea
  At dominant order in $\epsilon$ for each of the four pieces, this is the decomposition of Eq.~(\ref{deltaQ}):
\bsube
\bea 
    {_{I}\gamma}_{\alpha\beta} &
=& \left.{m}\ \frac\partial{\partial{m}}\gamma_{\alpha\beta} ({m},e)\right|_{({m},e)=(0,0)}\left[1+O(\epsilon)\right],\\
  {_{II}\gamma}_{\alpha\beta} &=& \left. e\ \frac\partial{\partial e} \gamma_{\alpha\beta}({m},e) \right|_{ ({m},e)=(0,0)}\left[1+O(\epsilon)\right],\\
   {_{I}A}_{\alpha} &=& \left. e\ \frac\partial{\partial e} A_\alpha ({m},e) \right|_{ ({m},e)=(0,0)}\left[1+O(\epsilon)\right],\\
   {_{II}A}_{\alpha} &=& \left.{m}\ \frac\partial{\partial{m}}A_\alpha ({m},e)\right|_{({m},e)=(0,0)}\left[1+O(\epsilon)\right].
\eea
\esube 

We can quickly find the short-distance (Hadamard) expansion of the 
solutions to Eqs.~(\ref{PertIG}) and (\ref{PertIE}), because their forms 
are nearly identical, respectively, to the equations governing the gravitational 
perturbation due to a massive particle with no charge, and to the electromagnetic 
perturbation due to a charged particle whose gravitational perturbation can 
be neglected.  Eq.~(\ref{PertIE}) is in fact the electromagnetic perturbation 
equation of a spacetime with no background electromagnetic field, but with the 
present background metric; its formal solutions are reviewed by Poisson {\em et al.} \cite{Poisson} 
and by the present authors in Paper I.   
Eq.~(\ref{PertIG}) differs from the equation governing the metric perturbation 
of a point mass in a vacuum spacetime only by the substitution 
$R\indices{_{\alpha}^{\gamma}_{\beta}^{\delta}} \rightarrow \hat{\Omega}\indices{_{\alpha}^{\gamma}_{\beta}^{\delta}}$. 
As seen in Appendix I, the Hadamard expansion of the field $_I\gamma_{\alpha\beta}$ differs only by the same substitution 
from the formal expansion found in Paper I for accelerated motion in a vacuum background.  

  The RNC expansions of solutions ${_{I}\gamma}_{\alpha\beta}$ and ${_{I}A}_{\alpha}$ to Eqs.~(\ref{PertIG}) and (\ref{PertIE}) are given by 

\bea
\frac{1}{{m}}\ {_{I}\gamma_{{{\alpha}}{{\beta}}}}&=&\frac{4u_{\alpha}u_{\beta}-8u_{(\alpha}a_{\beta)}u_{\gamma}x^{\gamma}}{\sqrt{S}} +4u_{{\mu}} u_{{\nu}} \hat{\Omega}\indices{_{(\alpha}^{\mu}_{\beta)}^{\nu}}\sqrt{S}
\nonumber\\
&+&\frac{4x^{{\mu}}x^{{\nu}}}{\sqrt{S}}\left[(a_{{{\alpha}}}a_{{\beta}} +\dot{a}_{({{\alpha}}}u_{{{\beta}})})(q_{{{\mu}}{{\nu}}}+u_{{\mu}} u_{{\nu}})+2a_{({{\alpha}}}u_{{{\beta}})}a_{{\mu}}u_{{\nu}} -\frac{u_{({{\alpha}}}R_{{{\beta}}){{\epsilon}}\gamma{{\sigma}}} u^{\gamma}}{3}(\delta\indices{^{{\epsilon}}_{{\mu}}}\delta\indices{^{{\sigma}}_{{\nu}}}+u^{{\epsilon}}\delta\indices{^{{\sigma}}_{{\mu}}}u_{{\nu}})
\right]\nonumber\\&+&O(\epsilon^2).
\label{Igamma}
\eea
and
\bea
\frac1e\ {}_I A_{{\alpha}}&=&\frac{u_\alpha-a_\alpha u_{\beta}x^{\beta}}{\sqrt{S}}
+
\frac{(u_\alpha R_{{{\gamma}}{{\delta}}}-2u^{{\beta}} R_{\alpha(\gamma\delta)\beta})}{12\sqrt{S}}\left[\delta\indices{^{{\gamma}}_{{\mu}}}\delta\indices{^{{\delta}}_{{\nu}}} +u^{{\gamma}}u^{{\delta}}(q_{{{\mu}}{{\nu}}}+u_{{{\mu}}} u_{{\nu}})+2u^{{\gamma}}\delta\indices{^{{\delta}}_{{\nu}}}u_{{\mu}}\right]x^{{\mu}}x^{{\nu}}\nonumber\\
&+&
\frac{\left[2a_\alpha u_{{\mu}} a_{{\nu}} +\dot{a}_\alpha (q_{{{\mu}}{{\nu}}}+u_{{\mu}} u_{{\nu}})\right]x^{{\mu}} x^{{\nu}}}{2\sqrt{S}}+\frac{6 R_{\alpha\beta}u^\beta-u_\alpha R}{12}\sqrt{S}+O(\epsilon^2),
\nonumber\\
\label{IA}
\eea
where $2S=q_{\alpha\beta}y^{\alpha}_{ret}y^{\beta}_{ret}+q_{\alpha\beta}y^{\alpha}_{adv}y^{\beta}_{adv},$ where the vector $y^\alpha$ is tangent to the unique null geodesic from the trajectory at $z(\tau_{ret})$ or $z(\tau_{adv})$ to the field point $x$.


   In Eqs.~(\ref{PertIIG}) and (\ref{PertIIE}) for $_{II}\gamma_{\alpha\beta}$ and 
$_{II} A_\alpha$, the left sides involve the same linear operators as 
those of Eqs.~(\ref{PertIG}) and (\ref{PertIE}).  The right sides 
are constructed not only from the solutions we have just obtained for 
$_I\gamma_{\alpha\beta}$ and $_I A_\alpha$ but also from the fields 
$_{II}\gamma_{\alpha\beta}$ and $_{II} A_\alpha$ themselves.  
We can obtain local solutions iteratively, noting that each 
solution is higher order in $\epsilon$ than its source.  In particular, 
the leading terms in $_I\gamma_{\alpha\beta}$ and $_I A_\alpha$ proportional to 
$1/\sqrt{S_0}$ give dominant terms in ${_{II}\gamma}_{\alpha\beta}$ and 
${_{II}A}_{\alpha}$ of subleading order, $O(\epsilon^0)$. 
The first iteration then uses on the right side the leading terms in 
$_I\gamma_{\alpha\beta}$ and $_I A_\alpha$: 
\bea
\Box\ {_{II}\gamma}_{\alpha\beta}+O(\epsilon^0)
   &=&-16e\Lambda\indices{_{\alpha\beta}^{\gamma\delta}} u_\delta|_{x=z(0)}\partial_{\gamma} 
		\left(\frac1{\sqrt{S_0}}\right)
\label{GammaIILeadingDEQ}\\
\Box\ {_{II}A}_{\alpha}+O(\epsilon^0)
  &=&-8{m} \Lambda\indices{_{\gamma\delta\alpha}^\beta}u^\gamma u^\delta|_{x=z(0)}
	\partial_\beta \left(\frac{1}{\sqrt{S_0}}\right),
\label{AIILeadingDEQ}
\eea 
where $S_0=q_{\alpha\beta}x^{\alpha}x^{\beta}$. 

Solving Eqs.~(\ref{AIILeadingDEQ}) and (\ref{GammaIILeadingDEQ}) as RNC expansions, we find
\bea
 {_{II}\gamma}_{\alpha\beta}&=&-8 e\frac{ u_{\delta}\Lambda\indices{_{\alpha\beta}^{\gamma\delta}}q_{\gamma\epsilon}x^{\epsilon}}{\sqrt{S_0}}+O(\epsilon)\nonumber\\
&=&-2{m}\frac{x^\gamma}{\sqrt{S_0}}\left(2 a_{(\alpha}\eta_{\beta)\gamma}-2{\frac e{m}} u_{(\beta}F_{\alpha)\gamma}- \eta_{\alpha\beta}a_{\gamma}\right)+O(\epsilon),
\label{GammaIILeading}
\\
{_{II}A}_{\alpha}&=&-4{m} \frac{u_{\gamma}u_{\delta}\Lambda\indices{^{\gamma\delta}_{\alpha\beta}}q^{\beta}_{\epsilon} x^{\epsilon}}{\sqrt{S_0}}+O(\epsilon)\nonumber\\
&=&-{{m}\over\sqrt{S_0}}\left[F_{\alpha\beta}
     +{{m}\over e} (a_\alpha u_\beta - 2 u_\alpha
a_\beta)\right]x^\beta+O(\epsilon).
\label{AIILeading}
\eea
Here and from now on, when the symbols $a_\alpha$, $u^\alpha$, $q_{\alpha\beta}$ and  
$F_{\alpha\beta}$ appear without explicit $x$ dependence, they denote 
the values of the corresponding quantities at the position $z(0)$ of the particle.

This first iteration is already enough for our principal results:  
The singular part of the self-force at leading and subleading order and, 
in particular, its contribution to the renormalized mass are unchanged by
the gravitational-electromagnetic coupling. The result is due to a remarkable 
cancellation of the contributions to the self force from the two mixed terms. 
That is, the contributions proportional to $em$ in the electromagnetic 
and gravitational parts of the self-force are equal and opposite. 
To see this, we compute the force using    
\bea
f_{\alpha}^{EM}&=& e\delta F_{\alpha\beta}u^\beta = \left(\delta_{\alpha}^{\beta}u^{\delta}-\delta_{\alpha}^{\delta}u^{\beta}\right)\partial_{\beta}\delta A_{\delta},
\nonumber\\
f_{\alpha}^{GR}
  &=& - {m} q_\alpha^\delta (\nabla_\beta h_{\gamma\delta}-\frac12\nabla_\delta h_{\beta\gamma} ) u^\beta u^\gamma = 
\frac{{m}}{4}\left[q_{\alpha}^{\beta}\left(q^{\gamma\delta}+u^{\gamma}u^{\delta}\right)-4q_{\alpha}^{\delta}u^{\beta}u^{\gamma}\right]\nabla_{\beta}\gamma_{\gamma\delta}.
\label{ForceEquations}
\eea
Substituting $_{II}\gamma_{\alpha\beta}$ and $_{II} A_\alpha$ in Eqs.~(\ref{GammaIILeading}) and (\ref{AIILeading}) gives the contributions proportional 
to $em$, namely  
\bea
{_{II}f^{EM}}_{\alpha}&=&-e{m} u^\beta F_{\gamma\beta}\left(\frac{\delta^{\gamma}_{\alpha}}{\sqrt{S_0}}-\frac{q_{\alpha\delta}x^{\delta}x^{\gamma}}{S_0^{3/2}}\right)+O(\epsilon^0)\nonumber\\
&=&- {_{II}f^{GR}}_{\alpha}.
\label{GRmixedSL}
\eea
 Note that the angle-average of each contribution, 
\be
 \langle{_{II}f^{s=1,2}}_{\alpha}\rangle
	= \mp\frac23  e{m}  F_{\alpha\beta}u^\beta \frac 1{\sqrt{S_0}}
	= \mp\frac23 \frac{{m}^2}{\sqrt{S_0}} a_\alpha,
\ee 
is proportional to $a_\alpha$ and would contribute to the mass renormalization 
if the terms did not cancel. 

The sums $_I\gamma_{\alpha\beta}+_{II}\gamma_{\alpha\beta}$ and $_I\delta A_\alpha+_{II}\delta A_\alpha$
are the singular fields to $O(\epsilon^0)$:
\bea
\gamma_{\alpha\beta}^{sing}&=&4\frac{{m} u_{\alpha}u_{\beta}-2\left({m} a_{(\alpha}u_{\beta)}u_{\epsilon}+eu_{\delta}\Lambda\indices{_{\alpha\beta}^{\gamma\delta}}q_{\gamma\epsilon}\right)x^\epsilon}{\sqrt{S}}+O(\epsilon),
\label{GammaFullSL}
\\
\delta A_{\alpha}^{sing}&=&\frac{eu_{\alpha}-\left(e a_{\alpha}u_{\epsilon}+4{m} u_{\gamma}u_{\delta}\Lambda\indices{^{\gamma\delta}_{\alpha\beta}}q^{\beta}_{\epsilon}\right)x^{\epsilon}}{\sqrt{S}}+O(\epsilon).
\label{AFullSL}
\eea

    We will now continue the iteration to obtain an $O(\epsilon)$ contribution 
$_{II}\gamma_{\alpha\beta}$ and $_{II} A_\alpha$ by including on the 
right side of Eqs.~(\ref{PertIIG}) and (\ref{PertIIE}) their known expansions through $O(\epsilon^0)$.
We obtain in this way the $O(\epsilon)$ contribution to the singular fields $\gamma_{\alpha\beta}^{sing}$ and 
$\delta A_\alpha^{sing}$ up to a homogeneous solution to the flat-space wave equation of the form 
$P^{(2n)}(x)/S_0^{n-1/2}$, where $P^{(2n)}$ is a homogeneous polynomial of degree $2n$ in the
coordinates $\{x^\mu\}$.       
Substituting the expressions (\ref{GammaFullSL}) and (\ref{AFullSL}) for $\gamma_{\alpha\beta}$ and $\delta A_\alpha$ 
back into Eqs.~(\ref{PertIIG}) and (\ref{PertIIE}) respectively, we have
\bea
\square {_{II}\gamma}_{\alpha\beta}+2\hat{\Omega}\indices{_{\alpha}^{\gamma}_{\beta}^{\delta}}{_{II}\gamma}_{\gamma\delta}&=&-16\Lambda\indices{_{\alpha\beta}^{\gamma\delta}}(x)\partial_{\gamma} \left(\frac{e u_{\delta} +A_{\delta\epsilon}x^{\epsilon}+O(\epsilon^2)}{\sqrt{S}}\right),
\label{PertIIGSSL}\\
\square {_{II}A}_{\alpha}-R_{\alpha}^{\beta}{_{II}A}_{\beta}&=&-\nabla^{\beta}\left[\Lambda\indices{^{\gamma\delta}_{\alpha\beta}}(x)\left(\frac{8{m} u_{\gamma}u_{\delta}+2\gamma_{\gamma\delta\epsilon}x^{\epsilon}+O(\epsilon^2)}{\sqrt{S}}\right)\gamma_{\gamma\delta}\right].
\label{PertIIESSL}
\eea
where $A_{\alpha\beta}$ and $\gamma_{\alpha\beta\gamma}$ are defined by
\bea
A_{\alpha\beta}&:=&-e a_{\alpha}u_{\beta}-4 {m} \Lambda_{\gamma\delta\alpha\epsilon}u^{\gamma}u^{\delta}q^{\epsilon}_{\beta}, 
\label{Asub2}\\
\gamma_{\alpha\beta\gamma}&:=&-8\left({m} a_{(\alpha}u_{\beta)}u_{\gamma}
     +e \Lambda_{\alpha\beta\delta\epsilon}q^\delta_\gamma u^\epsilon\right).
\label{gammasub3}
\eea
 The RNC expansion of $\Lambda\indices{_{\gamma\delta}^{\beta\alpha}}(x)$ about $z(0)$
is given by
\be
\Lambda\indices{_{\gamma\delta}^{\beta\alpha}}(x) = \Lambda\indices{_{\gamma\delta}^{\beta\alpha}}|_{x=z(0)}+\Lambda\indices{_{\gamma\delta}^{\beta\alpha}_{\epsilon}}x^{\epsilon}+O(\epsilon^2),
\label{LambdaExpand}\ee
where 
\be
\Lambda\indices{_{\gamma\delta}^{\beta\alpha}_{\epsilon}}
=\partial_{\epsilon}\Lambda\indices{_{\gamma\delta}^{\beta\alpha}}|_{x=z(0)}
=\left(\partial_{\epsilon}F\indices{_{(\alpha}^{[\delta}}\delta\indices{_{\beta)}^{\gamma]}}-\frac{1}{4}\eta_{\alpha\beta}\partial_{\epsilon}F^{\gamma\delta}\right)_{x=z(0)}.
\label{LambdaExpand2}
\ee 

 Solving Eqs.~(\ref{PertIIGSSL}) and (\ref{PertIIESSL}) for ${_{II}\gamma}_{\alpha\beta}$ and ${_{II}A}_{\alpha}$ to $O(\epsilon)$) 
and adding the result to the expansions of ${_{I}\gamma}_{\alpha\beta}$ and ${_{II}A}_{\alpha}$,  we obtain the singular fields to sub-subleading order, namely
\bea
\gamma^{local}_{\alpha\beta}&=&\frac{4{m} u_{\alpha}u_{\beta}+\gamma_{\alpha\beta\epsilon}x^{\epsilon}}{\sqrt{S}}+\frac{4{m} x^{\gamma}x^{\delta}\left[\left(a_{\alpha}a_{\beta}+\dot{a}_{(\alpha}u_{\beta)}\right)\left(q_{\gamma\delta}+u_{\gamma}u_{\delta}\right)+2 a_{(\alpha}u_{\beta)}a_{\gamma}u_{\delta}\right]}{\sqrt{S}}\nonumber\\
&-&\frac{4{m}}{3}\frac{u_{(\alpha}R_{\beta)\epsilon\gamma\delta}q^{\epsilon}_{\lambda}u^{\gamma}x^{\delta}x^{\lambda}}{\sqrt{S}}
+4{m} u_{\gamma}u_{\delta}\hat{\Omega}\indices{_{\alpha}^{\gamma}_{\beta}^{\delta}}\sqrt{S} +4\Lambda\indices{_{\alpha\beta}^{\gamma\delta}}A_{\delta\epsilon}\left(u^{\epsilon}u_{\lambda}-\delta^{\epsilon}_{\lambda}\right)\left(\delta^{\lambda}_{\gamma}\sqrt{S}+\frac{q_{\gamma\mu}x^{\mu}x^{\lambda}}{\sqrt{S}}\right)\nonumber\\
&+&4 e u_{\delta}\Lambda\indices{_{\alpha\beta}^{\gamma\delta}_{\epsilon}}\left[\left(u^{\epsilon}u_{\lambda}-\delta^{\epsilon}_{\lambda}\right)\frac{q_{\gamma\mu}x^{\lambda}x^{\mu}}{\sqrt{S}}+q^{\epsilon}_{\gamma}\sqrt{S}\right]+O(\epsilon^2),
\label{gammaSing}
\eea
and
\bea
\delta A^{local}_{\alpha}&=&\frac{e u_{\alpha}+A_{\alpha\beta}x^{\beta}}{\sqrt{S}}+e\left(\frac{u_{\alpha}R_{\gamma\delta}-2u^{\beta}R_{\alpha(\gamma\delta)\beta}}{12\sqrt{S}}\right)\left[\delta^{\gamma}_{\mu}\delta^{\delta}_{\nu}+2u^{\gamma}u_{\mu}\delta^{\delta}_{\nu}+u^{\gamma}u^{\delta}\left(q_{\mu\nu}+u_{\mu}u_{\nu}\right)\right]x^{\mu}x^{\nu}
\nonumber\\
&+&\frac{e}{2}\left[\frac{2a_{\alpha}a_{\delta}u_{\gamma}+\dot{a}_{\alpha}(q_{\gamma\delta}
	+u_{\gamma}u_{\delta})}{\sqrt{S}}\right]x^{\gamma}x^{\delta}
	+\frac{e}{2}u^{\beta}R_{\alpha\beta}\sqrt{S}
	+4{m} \frac{\Lambda_{\gamma\delta\alpha\beta}u^\beta u^\gamma u^\delta u_\mu a_\nu x^\mu x^\nu}{\sqrt{S}}\nonumber\\
&+&\frac{4{m} u_{\gamma}u_{\delta}\Lambda\indices{^{\gamma\delta}_{\alpha\beta\epsilon}}+\Lambda\indices{^{\gamma\delta}_{\alpha\beta}}\gamma_{\gamma\delta\epsilon}}{2}\left(u^{\epsilon}u_{\lambda}-\delta^{\epsilon}_{\lambda}\right)\left(\eta^{\lambda\beta}\sqrt{S}+\frac{q^{\beta}_{\mu}x^{\lambda}x^{\mu}}{\sqrt{S}}\right)+O(\epsilon^2).
\label{ASing}
\eea

The fields $\gamma^{local}_{\alpha\beta}$ and $\delta A^{local}_{\alpha\beta}$ coincide with the singular fields at leading and subleading order.  We could now impose conditions on the Green's function 
analogous to those of Eqs.~(\ref{U2}) and (\ref{V2}) to complete the sub-subleading part of the 
singular field, and from it compute the sub-subleading piece of the singular force, but we are saved the trouble: As in Eqs.~(\ref{gammaSing}) and (\ref{ASing}) the sub-subleading terms are functions even in the coordinates $x^\mu$. Because the expressions for the self-force in Eqs.~(\ref{ForceEquations}) are proportional to the gradients of the potentials, they are 
odd in $x^\mu$ and will therefore vanish upon angle averaging. 
The remaining contributions to the self-force are at leading 
and and subleading order, 
$O(\epsilon^{-2})$ and $O(\epsilon^{-1})$, and we have
\bea
f_{\alpha}^{sing}&=&\left(e^2-{m}^2\right)\left[\frac{q_{\alpha\beta}x^{\beta}}{S_0^{3/2}}-\left[q_{\alpha\beta}a_{\gamma}\left(3\eta_{\epsilon\delta}-2q_{\epsilon\delta}\right)-a_{\alpha}q_{\gamma\delta}\eta_{\epsilon\beta}\right]\frac{x^{\gamma}x^{\delta}x^{\beta}x^{\epsilon}}{S_0^{5/2}}\right]\nonumber\\
&+&\frac{(4{m}^2-e^2)a_{\alpha}}{\sqrt{S_0}}+O(\epsilon^0).
\label{SingularForce}
\eea
The term $O(\epsilon^0)$ can be written as a seventh order polynomial in $x^\mu$ 
divided by $S_0^{7/2}$, manifestly odd in the RNCs.  This implies not only their 
angle average vanishes, but also that they do not contribute to the regularization 
parameters in mode-sum regularization.  

With the cancellation of the mixed terms (terms proportional to $em$) in the 
expression for the self-force, $f_\alpha^{sing}$ at subleading order is 
unaltered by the coupling of the electromagnetic and gravitational fields 
when it is written in terms of $g_{\alpha\beta}$, $u_\alpha$, $a_\alpha$ and the RNCs.
A charge $e$ moving with this acceleration in a geometry with this metric but with 
no background electromagnetic field has $f_\alpha^{sing}$ given by 
the part of the present $f_\alpha^{sing}$ that is proportional to $e^2$; and, 
as shown in Paper I, a mass $m$, again moving on the same accelerated 
trajectory but with non-gravitational interactions ignored, has an $f_\alpha^{sing}$
given by the terms proportional to ${m}^2$.  As we discuss in the 
next section only the leading and subleading regularization parameters 
$A_\alpha$ and $B_\alpha$ are requrired for mode-sum regularization, and 
they are each simply the sum of similarly independent electromagnetic 
and gravitational parameters.  

  In Appendix B, using the potentials obtained by Zimmerman and Poisson 
for a particle of scalar charge $q$ and mass $m$ moving in a 
scalarvac spacetime, we find that the analogous result holds.  Again 
to subleading order, there is no mixed contribution to the singular 
expression for the self-force; $f_\alpha^{sing}$ is at this order the 
sum of its purely gravitational and scalar terms; and the mode-sum 
regularization requires only  parameters 
$A_\alpha$ and $B_\alpha$ that are each the sum of independent 
gravitational and electromagnetic parameters.  

\section{Mode-sum Regularization}
\label{SectionIII}

In mode-sum regularization, one decomposes retarded and singular fields in angular 
harmonics associated with a given set of spherical coordinates.  The method is commonly used 
for black-hole spacetimes with Schwarzschild or Boyer-Lindquist coordinates, for example, 
but the formalism is valid for arbitrary spherical coordinates $(\hat x^\mu)=(t,r,\theta,\phi)$ 
related in the usual way to a smooth Cartesian chart $(t,x^1,x^2,x^3)$, for which 2-spheres 
of constant $t$ and $r$ are in the domain of the chart. 
We choose $t=0$ at the position $z(0)$ of the particle and denote its coordinates by 
$(0,r_0,\theta_0,\phi_0)$. 
Using the angular harmonic decomposition of the field perturbations, one similarly 
decomposes $f^{ret}_\alpha$ and $f^{sing}_\alpha$ as sums of of angular harmonics 
on a sphere through the particle and writes the renormalized self-force as 
the convergent sum
\be
f^{ren}_{\alpha}=\sum_{\ell=0}^{\infty} (f^{ret,\ell}_\alpha -f^{sing,\ell}_\alpha), 
\label{SFmodesum}
\ee
where
\be
 f^{ret/sing,\ell}_\alpha
   = \lim_{r\rightarrow r_0}\sum_{\texttt{m}=-\ell}^{\ell}\int d\Omega 
		f^{ret/sing}_{\alpha}(0,r,\theta_0,\phi_0) \bar{Y}_{\ell \texttt{m}}(\theta,\phi).
\ee
\
(Here $f^{ret,\ell }_{\alpha}$ and hence $f^{sing,\ell }_{\alpha}$ have leading terms whose sign 
depends on whether the $r\rightarrow 0$ limit is taken from $r<r_0$ or $r>r_0$.)
Because $f^{sing}_\alpha$ is defined only in a normal neighborhood of $z(0)$ one first arbitrarily 
extends it to a thickened 2-sphere, and $f^{sing}_{\alpha}$ then refers to this 
smooth extension.  As mentioned in Sect.~\ref{Introduction}, Paper I generalized to accelerated 
motion in generic spacetimes a result found by Barack and Ori \cite{BO00} for geodesic 
motion in Schwarzschild and Kerr geometries, that $f^{sing,\ell}$ has the 
form $A_\alpha L + B_\alpha + O(L^{-2})$, 
where $L=\ell+1/2$, $A_\alpha$ and $B_\alpha$ are independent of $\ell$, and  
the convergent sum of the $O(L^{-2})$ part of $f^{sing,\ell}$ vanishes in a Lorenz gauge.  

The singular fields, $\gamma^{sing}_{\alpha\beta}$ and $\delta A^{sing}_\alpha$, differ at subleading order from the values computed in Paper I for accelerated motion of an electric 
charge $e$ and of a point mass $m$. The difference in the expression for the singular part of the self-force, however, coincides at subleading order, with the sum of the electromagnetic and gravitational expressions in Paper I. As in Paper I, the terms of this order have the form 
of a polynomial of odd degree in the RNCs, divided by a half-integral power of $S_0$, 
and that fact, together with the form of the leading and subleading terms, implies 
Eq.~(\ref{eq:fsingAB}).  We then have the result 
\be
f^{ren}_{\alpha}=\sum_{\ell=0}^{\infty} [f^{ret,\ell}_\alpha -(A_\alpha L + B_\alpha)], 
\ee
where the values of $A_\alpha$ and $B_\alpha$ are just the sums 
\be
  A_\alpha = A^{GR}_\alpha+A^{EM}_\alpha, \qquad B_\alpha = B^{GR}_\alpha+B^{EM}_\alpha,
\ee 
of the gravitational and electromagnetic parameters obtained in Paper I from the 
perturbed gravitational and electromagnetic fields of accelerated 
charges and masses associated with a choice $(t,r,\theta,\phi)$ of spherical 
coordinates.  These are given in Eqs.~(C20) and (C21) of Paper I.  We display the explicit values of $A_{\alpha}$ and $B_{\alpha}$ in Appendix~\ref{parameters}.

\section{Conclusions} 
 
With the assumption that one can renormalize the self-force on a charged point mass moving 
in an electrovac spacetime by a combination of angle-average and mass renormalization, 
we show that the renormalization can be done as if the equations for the perturbed 
electromagnetic and gravitational fields were decoupled.  We showed that, as in the 
case of uncoupled fields, the renormalized self-force can by obtained by a mode-sum 
regularization that involves only the leading and subleading 
regularization parameters $A_\alpha$ and $B_\alpha$; and each of these parameters 
is a sum of their previously calculated values for purely electromagnetic and gravitational
perturbations from an accelerated point charge and a point mass, respectively. 
The renormalization we obtain is also equivalent to that found in independent work 
by Zimmerman and Poisson.  The perturbed fields are, of course, coupled, and the 
renormalized self-force has mixed terms, 
terms proportional to $em$ as well as to $e^2$ and ${m}^2$. 

What has not yet been done is to extend to the coupled fields we consider here 
the underlying justification based on variants of matched asymptotic expansion 
for electromagnetic or gravitational interactions alone.  The equivalence of 
the several renormalization prescriptions that we and Poisson and Zimmerman 
have considered, however, gives us confidence that no surprise will 
emerge.  

Finally, the results obtained here may be useful in studying the problem of overcharging a black hole, 
with charged particles moving in a background Riessner-Nordstrom spacetime.  

\begin{acknowledgments}
This work was supported in part by NSF Grants PHY 1001515, PHY 1307429, and PHY 0970074. 
We thank Eric Poisson and Peter Zimmerman for a comparison of their results and ours on the fields 
and renormalization of charged, massive particles in electrovac spacetimes.  They also 
kindly allowed us to use their results for a scalar particle to check the decoupling 
of scalar and gravitational contributions to the expression for the self-force at 
subleading order. 
\end{acknowledgments}
\appendix
\section{Gravitational Green's function in a non-vacuum spacetime}
\label{Hadamard}
We will make extensive use of the treatment found in \cite{Poisson}. The goal is to find the Green's function $G\indices{^{\alpha\beta}_{\gamma'\delta'}}(x,x')$, where $x$ and $x'$ are two arbitrary points in a convex normal neighborhood $C$, and unprimed and primed indices 
are tensor indices at $x$ and $x'$, respectively. When we apply this to solve for ${_I\gamma}_{\alpha\beta}$ in Eq.~(\ref{Igamma}), we set $x'=z(0)$. We consider the purely gravitational Green's function, the solution to
\be
\Box G\indices{^{\alpha\beta}_{\gamma'\delta'}}(x,x')+2\hat{\Omega}\indices{^{\alpha}_{\mu}^{\beta}_{\nu}}G\indices{^{\mu\nu}_{\gamma'\delta'}}(x,x')=
-4\pi g\indices{^{(\alpha}_{\gamma'}}(x,x')g\indices{^{\beta)}_{\delta'}}(x,x')\delta^{(4)}(x,x'),
\label{GravityGreen'sFunction}
\ee
where  $g\indices{^{\alpha}_{\gamma'}}(x,x')$ is the bivector of parallel transport, taking a vector, $v^{\gamma'}(x')$, defined at $x'$ and parallel transporting it along the unique geodesic connecting $x$ and $x'$, resulting in $v^{\alpha}(x,x')=g\indices{^{\alpha}_{\gamma'}}v^{\gamma'}(x')$.

 The retarded and advanced Green's functions $G\indices{^{\alpha\beta}_{\gamma'\delta'}_{\pm}}(x,x')$ have the form,
\be
G\indices{^{\alpha\beta}_{\gamma'\delta'}_{\pm}}(x,x')=U\indices{^{\alpha\beta}_{\gamma'\delta'}}(x,x')\delta_{\pm}(\sigma)+V\indices{^{\alpha\beta}_{\gamma'\delta'}}(x,x')\theta_{\pm}(-\sigma),
\label{Ansatz}
\ee
where the distributions $\delta_{\pm}$ and $\theta_{\pm}$ are defined in Section 13 of \cite{Poisson}, and $\sigma$ is Synge's world function.  Substituting Eq.~(\ref{Ansatz}) into the left hand side of Eq.~(\ref{GravityGreen'sFunction}), we find (with the argument $(x,x')$ of bitensors suppressed)
\bea
\Box G\indices{^{\alpha\beta}_{\gamma'\delta'}}+2\hat{\Omega}\indices{^{\alpha}_{\mu}^{\beta}_{\nu}}G\indices{^{\mu\nu}_{\gamma'\delta'}}&=&
-4\pi U\indices{^{\alpha\beta}_{\gamma'\delta'}}\delta^{(4)}(x,x')+\delta'_{\pm}(\sigma)\left(2U\indices{^{\alpha\beta}_{\gamma'\delta';\gamma}}\sigma^{\gamma}+(\sigma\indices{^{\gamma}_{\gamma}}-4)U\indices{^{\alpha\beta}_{\gamma'\delta'}}\right)\nonumber\\
&+&\delta_{\pm}(\sigma)\left(-2V\indices{^{\alpha\beta}_{\gamma'\delta';\gamma}}\sigma^{\gamma}+(2-\sigma\indices{^{\gamma}_{\gamma}})V\indices{^{\alpha\beta}_{\gamma'\delta'}}+(\Box U\indices{^{\alpha\beta}_{\gamma'\delta'}}+2\hat{\Omega}\indices{^{\alpha}_{\mu}^{\beta}_{\nu}}U\indices{^{\mu\nu}_{\gamma'\delta'}})\right)\nonumber\\
&+& \theta_{\pm}(-\sigma)\left(\Box V\indices{^{\alpha\beta}_{\gamma'\delta'}}+2\hat{\Omega}\indices{^{\alpha}_{\mu}^{\beta}_{\nu}}V\indices{^{\mu\nu}_{\gamma'\delta'}}\right)\nonumber\\
&=&-4\pi g\indices{^{(\alpha}_{\gamma'}}g\indices{^{\beta)}_{\delta'}}\delta^{(4)}(x,x').
\label{AppendixMainEquation}
\eea
In comparing this to the corresponding (unnumbered) equation in \cite{Poisson} (between Eq. 16.7 and 16.8),
it is clear that the only difference is that the tensor $R\indices{^{\alpha}_{\gamma}^{\beta}_{\delta}}$ is replaced here by $\hat{\Omega}\indices{^{\alpha}_{\gamma}^{\beta}_{\delta}}$. 
Following the same technique used in \cite{Poisson}, we require that the coefficients of $\delta'_{\pm}(\sigma)$, $\delta_{\pm}(\sigma)$, and $\theta_{\pm}(\sigma)$ separately 
vanish. We thereby find,
\bea
U\indices{^{\alpha\beta}_{\gamma'\delta'}}(x,x') 
	&=& g\indices{^{(\alpha}_{\gamma'}}(x,x')g\indices{^{\beta)}_{\delta'}}(x,x')\Delta^{1/2}(x,x')\nonumber\\
   &=& g\indices{^{(\alpha}_{\gamma'}}(x,x')g\indices{^{\beta)}_{\delta'}}(x,x')\left(1+\frac{1}{12}R_{\gamma'\delta'}\sigma^{\gamma'}\sigma^{\delta'}+O(\epsilon^3)\right),
\label{U2}
\eea
and
\be
V\indices{^{\alpha\beta}_{\gamma'\delta'}}(x',x') = \frac{\delta\indices{^{(\alpha'}_{\gamma'}}\delta\indices{^{\beta')}_{\delta'}} R(x')}{12}+ \hat{\Omega}\indices{^{\alpha'}_{(\gamma'}^{\beta'}_{\delta')}}(x').
\label{V2}
\ee
Since $R=0$ for electrovac, the only difference between the Hadamard expansions of a point mass in vacuum and ${_{I}\gamma}_{\alpha\beta}$ is in the bitensor $V\indices{^{\alpha\beta}_{\gamma'\delta'}}$, where instead of the Riemann tensor we have $\hat{\Omega}\indices{^{\alpha'}_{(\gamma'}^{\beta'}_{\delta')}}$.
\section{Decoupling in renormalization of a massive scalar charge.}

Using work on a massive scalar charge by Zimmerman and Poisson \cite{pz14} (ZP), 
we verify here that there is no cross-term at subleading order in the singular 
expression for the self-force of a massive particle with scalar charge 
moving in a background scalarvac spacetime.  The result implies that, 
as in the case of a point charge in an electrovac spacetime, the 
renormalized mass is obtained by subtracting 
(1) the scalar-field contribution from a point charge moving along 
an accelerated trajectory and (2) the purely gravitational contribution 
from a point mass moving along the same trajectory.  In a mode-sum regularization, 
the regularization parameters are then sums of their purely scalar and 
gravitational values.  

Subleading terms in $f^{sing}_\alpha$ proportional to $q\frak{m}$ arise from 
terms of order $\epsilon^0$ in $\Phi^{sing}$ that are proportional to ${m}$ 
and from terms of order $\epsilon^0$ in $\gamma^{sing}_{\alpha\beta}$ that are 
proportional to $q$. 
We consider first the contribution to the self-force from $\Phi^{sing}$. 
From Eq.~(6.19) of \cite{pz14},
written in terms of our RNCs with origin at $z(0)$, we have
\be
   \Phi^{sing} = \frac{1}{\sqrt{S_0}}[\gamma_1 U +u_{\alpha}x^{\alpha}\gamma_2 \dot U + O(\epsilon^2)],
\ee
where $\gamma_1$ and $\gamma_2$ are independent of the perturbed fields, with 
$\gamma_1[z(0)] = \gamma_2[z(0)] = 1$. From Eq.~(7.25) of ZP, $U$ and $\dot U$ 
have no terms proportional to $\frak{m}$, implying that there is no $q\frak{m}$ contribution to $f^{sing}_\alpha$ from $\Phi^{sing}$ at subleading order, $O(\epsilon^{-1})$.  

We turn next to the contribution from $\gamma^{sing}_{\alpha\beta}$.  The 
symbol $\hat r$ in ZP is $\hat r = u_{\epsilon}x^{\epsilon}$.  Again from Eq.~(6.19),
$$ 
  \gamma_{sing}^{\alpha\beta} =  {1\over\rho}[\gamma_1 U^{\alpha\beta} 
		+u_{\epsilon}x^{\epsilon} \gamma_2 \dot U^{\alpha\beta}  + O(\epsilon^2)],
$$
with (7.25) giving no term in $U^{\alpha\beta}$ proportional to $q$ and, in 
$\dot U^{\alpha\beta}$, the single term 
$$
q{\partial\over\partial q} \dot U^{\alpha\beta} = -4 q\dot\Phi u^\alpha u^\beta.
$$
From (6.21), $\gamma_2 = 1+O(\epsilon)$, and 
the single term proportional to $q$ in $\gamma^{\alpha\beta}$ is then 
$$
q{\partial\over\partial q} \gamma^{\alpha\beta} =- 4\frac{u_{\gamma}x^{\gamma}}{\sqrt{S_0}} q\dot\Phi u^\alpha u^\beta.
$$

The contribution of this term to the self-force at subleading order is then
\be
{{m}\over 4}\left.\left[q_i^\beta(q^{\gamma\delta}+u^\gamma u^\delta) -4 q_i^\gamma u^\beta u^\delta\right]
	\nabla_\beta\,\gamma_{\gamma\delta}\right|_{t=0} = {m} q\dot{\Phi}\left[\frac{u_{\beta}x^{\beta}q_{\alpha\gamma}x^{\gamma}}{S_0^{3/2}}\right]_{t=0}= 0, 
\ee
using $u_i = 0$.  
We conclude that there is no contribution to the self force through subleading order
that is proportional to $q{m}$. 
  
\section{Integrals for the B term}
\label{parameters}

We give here the explicit forms of the regularization parameters $A_\alpha$ and $B_\alpha$ associated 
with arbitrary spherical coordinates $t,r,\theta,\phi$, with $(t_0,r_0,\theta_0,\phi_0)$ the coordinates 
of the position of the particle.  
The value of $A_\alpha$ is obtained as a limit $r\rightarrow r_0$ from $r<r_0$ or $r>r_0$, and it is given by 
\be
A_{\alpha\pm}=\mp(e^2-{m}^2)\sin\theta_0
\frac{q_{\alpha r}-q_{\alpha\theta}q_{\theta r}/q_{\theta\theta}-q_{\alpha\phi}q_{\phi r}/q_{\phi\phi}}{\left(q_{\phi\phi}q_{\theta\theta}q_{rr}-q_{\phi\phi}q^2_{\theta r}-q_{\theta\theta}q^2_{\phi r}\right)^{1/2}}.
\label{Aterm}
\ee

In writing the components of $B_\alpha$, it is helpful to define a tensor $c^{\alpha}_{\beta\gamma}$ at $z(0)$, 
whose only nonvanishing components are $c^{\theta}_{\phi\phi}=4^{-1}\sin(2\theta_0)$ and $c^{\phi}_{\theta\phi}=c^{\phi}_{\phi\theta}=-2^{-1}\cot\theta_0$. 
Then 
\be
B_{\alpha}=\frac{P_{\alpha\beta\gamma\delta\epsilon}}{2\pi}I^{\beta\gamma\delta\epsilon},
\ee
where 
\bea
P_{\alpha\beta\gamma\delta\epsilon}&=&(e^2-{m}^2)\left[q_{\alpha\beta}a_{\gamma}\left(3g_{\epsilon\delta}-2q_{\epsilon\delta}\right)-a_{\alpha}q_{\gamma\delta}g_{\epsilon\beta}+\left(3q_{\beta\lambda}q_{\alpha\epsilon}-q_{\alpha\lambda}q_{\beta\epsilon}\right)c^{\lambda}_{\gamma\delta}\right]\nonumber\\
&+&(4{m}^2-e^2)a_{\alpha}q_{\beta\gamma}q_{\delta\epsilon}
\label{Ptilde}
\eea
and the components $I^{\alpha\beta\gamma\delta}$ are zero unless all indices are either $\theta$ or $\phi$.
The nonzero values of $I^{\alpha\beta\gamma\delta}$ are all proportional to functions of the quantities 
\be
\alpha = \sin^2\theta_0\ q_{\theta\theta}/q_{\phi\phi}-1 \mbox{ and } \beta=2\sin\theta_0\ q_{\theta\phi}/q_{\phi\phi}.
\ee
To display them, we follow Barack \cite{barack09}, defining an integer $N$, $0\leq N\leq 4$, whose value is the number of times the index $\phi$ occurs in the set $\{\alpha,\beta,\gamma,\delta\}$: $N= \delta_\alpha^\phi+ \delta_\beta^\phi+\delta_\gamma^\phi+\delta_\delta^\phi$.
Then, when each index of $I^{\alpha\beta\gamma\delta}$ is $\theta$ or $\phi$, 
\be
I^{\alpha\beta\gamma\delta} =\frac{(\sin\theta_0)^{5-N}}{(\alpha^2+{\beta}^2)^2(4\alpha+4-{\beta}^2)^{3/2}(Q/2)^{1/2}}\left[QI^{(N)}_K \hat{K}(\omega)+I_{E}^{(N)}\hat{E}(\omega)\right],
\ee
where
\be
Q=\alpha+2-(\alpha^2+{\beta}^2)^{1/2} \mbox{ and } \omega=\frac{2(\alpha^2+{\beta}^2)^{1/2}}{\alpha+2+(\alpha^2+{\beta}^2)^{1/2}}. 
\ee

The various integrals for the $B$ term, computed first in \cite{bo03} are given below:
\begin{eqnarray}
I^{(0)}_{K}=4\left[12\alpha^3+\alpha^2(8-3{\beta}^2)-4\alpha{\beta}^2+{\beta}^2({\beta}^2-8)\right],\nonumber\\
I^{(0)}_{E}=-16\left[8\alpha^3+\alpha^2(4-7{\beta}^2)+\alpha{\beta}^2({\beta}^2-4)-{\beta}^2({\beta}^2+4)\right],
\label{I(0)}
\end{eqnarray}
\begin{eqnarray}
I^{(1)}_{K}=8{\beta}\left[9\alpha^2-2\alpha({\beta}^2-4)+{\beta}^2\right],\nonumber\\
I^{(1)}_{E}=-4{\beta}\left[12\alpha^3-\alpha^2({\beta}^2-52)+\alpha(32-12{\beta}^2)+{\beta}^2(3{\beta}^2+4)\right],
\label{I(1)}
\end{eqnarray}
\begin{eqnarray}
I^{(2)}_{K}=-4\left[8\alpha^3-\alpha^2({\beta}^2-8)-8\alpha{\beta}^2+{\beta}^2(3{\beta}^2-8)\right],
\nonumber\\
I^{(2)}_{E}=8\left[4\alpha^4+\alpha^3({\beta}^2+12)+\alpha({\beta}^2-4)(3{\beta}^2-2\alpha)+2{\beta}^2(3{\beta}^2-4)\right],
\label{I(2)}
\end{eqnarray}
\begin{eqnarray}
I^{(3)}_{K}=8{\beta}\left[\alpha^3-7\alpha^2+\alpha(3{\beta}^2-8)+{\beta}^2\right],\nonumber\\
I^{(3)}_{E}=-4{\beta}\left[8\alpha^4-4\alpha^3+\alpha^2(15{\beta}^2-44)+4\alpha(5{\beta}^2-8)+{\beta}^2(3{\beta}^2+4)\right],
\label{I(3)}
\end{eqnarray}
\begin{eqnarray}
I^{(4)}_{K}=-4\left[4\alpha^4-4\alpha^3+\alpha^2(7{\beta}^2-8)+12\alpha{\beta}^2-{\beta}^2({\beta}^2-8)\right],\nonumber\\
I^{(4)}_{E}=16\left[4\alpha^5+4\alpha^4+\alpha^3(7{\beta}^2-4)+\alpha^2(11{\beta}^2-4)+(2\alpha+1){\beta}^2({\beta}^2+4)\right].
\label{I(4)}
\end{eqnarray}
\bibliography{Biblio}
\end{document}